\documentstyle[preprint,aps,pre,psfig]{revtex}
\begin{document}
\title{Elastic moduli, dislocation core energy and melting of hard disks 
in two dimensions} 

\author{Surajit Sengupta$^1$\thanks{{\it on leave from}: Material Science
Division, Indira Gandhi Centre for Atomic Research, Kalpakkam 603102, India},
Peter Nielaba$^2$, and K. Binder$^1$}

\address{$^1$ Institut f\"ur Physik, Johannes Gutenberg Universit\"at Mainz,
55099, Mainz, Germany \\ $^2$ Universit\"at Konstanz, Fakult\"at f\"ur
Physik, Fach M 691, 78457, Konstanz, Germany \\}

\date{\today}

\maketitle

\begin{abstract}
Elastic moduli and dislocation core energy of 
the triangular solid of hard disks of diameter $\sigma$
are obtained in the limit of vanishing dislocation- antidislocation pair 
density, from Monte Carlo simulations which incorporates a 
constraint, namely that all moves altering the local connectivity
away from that of the ideal triangular lattice are rejected.
In this limit, we show that the solid is stable against all other 
fluctuations at least upto densities as low as $\rho \sigma^2 = 0.88$.
Our system does not show any phase transition so diverging correlation 
lengths leading to finite size effects and slow relaxations do not exist. 
The dislocation pair formation probability is estimated from the fraction of
moves rejected due to  the constraint which yields,
in turn, the core energy $E_c$ and the (bare) dislocation fugacity $y$. 
Using these quantities, we check the relative validity of first 
order and Kosterlitz~-Thouless~-Halperin~-Nelson~-Young (KTHNY) melting 
scenarios and obtain numerical estimates 
of the typical expected transition densities and pressures. We conclude that a 
KTHNY transition from the 
solid to a hexatic phase preempts the 
solid to liquid first order transition in this system albeit by a very small 
margin, easily masked by crossover effects in unconstrained 
``brute~- force'' simulations with small number of particles. 
\end{abstract}


\section{Introduction}
One of the first continuous systems to be studied by 
computer simulations\cite{metro,aldwain} is the system of hard disks
interacting with the two body potential,
\begin{eqnarray}
V(r) & = & \infty, r \le \sigma \\ \nonumber
     & = & 0,      r > \sigma
\end{eqnarray}
where, $\sigma$ (taken to be $1$ in the rest of the paper) the hard 
disk diameter, sets the length scale for the system and the energy 
scale is set by $k_B T = 1$. Despite its simplicity\cite{simple}, this 
system was 
shown to undergo a phase transition from solid to liquid as the 
density $\rho$ was decreased. The nature of this phase transition, 
however, is still being debated.   
Early simulations~\cite{aldwain,wwwood}
always found strong first order transitions.
As computational power increased the observed strength of
the first order transition progressively decreased!
Using sophisticated techniques Lee and Strandburg~\cite{hdstrand} and 
Zollweg and Chester~\cite{zolche} found evidence for, at best, a weak first
order transition. A first order transition has also been predicted 
by theoretical approaches based on density functional theory\cite{dft}.
On the other hand, recent simulations of hard disks~\cite{Jas} by Jaster, 
using as many as $N = 65536$
particles find evidence for a continuous, Kosterlitz -Thouless -Halperin
-Nelson -Young (KTHNY) transition\cite{KTHNY} from liquid to a hexatic
phase, with orientational but no translational order, at
$\rho = 0.899$. Nothing could be ascertained, however, about the expected 
hexatic to the crystalline solid transition at higher densities because 
the computations became prohibitively expensive. The solid to hexatic 
melting transition was estimated to occur at a density $\rho_c \geq .91$.  
A priori, it is difficult to assess why 
various simulations give contradicting results concerning the order of the 
transition. In this paper we take an approach, complementary to 
Jaster's, and investigate the melting transition of the solid 
phase. We show that the hard disk solid is unstable to perturbations 
which attempt to produce free dislocations leading to a solid $\to$ hexatic 
transition in accordance with KTHNY theory\cite{KTHNY}. Though this has been 
attempted in the past\cite{WB,bates}, numerical difficulties, especially 
with regard to equilibration of defect degrees of freedom, makes this task 
highly challenging. We also show that this 
transition lies close to a, first order, solid to liquid melting line. We 
calculate quantitatively the relative positions of the first order and the 
KTHNY transitions in the parameter space for this system and explain 
why earlier simulations failed to arrive at a consensus. 

The coarse grained density of a crystalline solid can be expanded as,
$\rho({\bf r}) = \sum_{\bf G} \rho_{\bf G} e^{i {\bf G}\cdot{\bf r}}$, where
({\bf G}) is a reciprocal lattice vector. 
The order parameters $\rho_{\bf G}$ are {\em complex},
$\rho_{\bf G} = |\rho_{\bf G}|e^{i {\bf u}\cdot{\bf G}},$ and
the displacement vector ${\bf u}$ is the deviation of an atom from
the nearest perfect lattice point ${\bf R}$.
If fluctuations of the amplitude of $\rho_{\bf G}$ can be neglected then 
a solid can be described in terms of ${\bf u }$ alone ---- the fundamental 
assumption of elasticity theory.
The elastic Hamiltonian for hard disks is given by,
\begin{equation}
F = -P \epsilon_{+} + B/2 \epsilon_{+}^2 +
(\mu+P)(\epsilon_{-}/2 + 2 \epsilon_{xy}),
\end{equation}
where $B$ is the
bulk modulus. The quantity $\mu_{eff} = \mu + P$ is the ``effective'' shear 
modulus
(the slope of the shear stress vs shear strain curve) and $P$ is the 
pressure. The hard disk solid, being a purely repulsive system,
is always under a uniform hydrostatic
pressure $P(\rho)$ at any density $\rho$. The Lagrangian elastic strains are 
defined as,
\begin{equation}
\epsilon_{ij} = \frac{1}{2} \left( \frac{\partial u_i}{\partial R_j} +
\frac{\partial u_j}{\partial R_i}+\frac{\partial u_i}{\partial R_k}
\frac{\partial u_k}{\partial R_j} \right),
\end{equation}
where the indices $i,j$ go over $x$ and $y$ and finally,
$
\epsilon_{+} = \epsilon_{xx}+\epsilon_{yy},$ and
$
\epsilon_{-} = \epsilon_{xx}-\epsilon_{yy}$.

In general a solid possesses two 
types of excitations, ``smooth'' phonons and ``singular'' dislocations,
respectively. 
Long wavelength phonons inhibit long range order in 2-d solids 
so that the intensity of a Bragg reflection peak $I_G \sim e^{-2W_G}$, 
where the Debye Waller factor $ W_G \sim G^2 a^{2-d}/(d-2)$ 
($a$ is the lattice parameter and $d$ the number of spatial dimensions) 
diverges and order parameter correlations decay algebraically ----
an example of Quasi Long Ranged Order (QLRO). We know that singular 
excitations, like dislocations, can drive a QLRO $\to$ disorder 
transition (where correlations decay exponentially). This situation
has been analysed by the KTHNY theory~\cite{KTHNY}. 

The KTHNY- theory~\cite{KTHNY} is presented usually
for a 2-d triangular solid under {\em zero external stress}.
It is shown that the dimensionless Young's modulus 
of a two-dimensional solid,
$$ K = \frac{8}{\sqrt 3 \rho} \frac{\mu}{\lbrace 1+\mu/(\lambda+\mu) 
\rbrace},$$ 
where $\mu$ and $\lambda$ are the Lam\'e constants,
depends on the fugacity of dislocation pairs, 
$y=\exp (-E_c)$, where $E_c$ is the core energy of the dislocation,
and the ``coarse~-graining'' length scale $l$. This dependence is
expressed in the form of the following coupled differential equations 
(the recursion relations) for the renormalization of $K$ and $y$:
\begin{eqnarray}
\frac{\partial K}{\partial l} & = & 3 \pi y^2 e^{\frac {K}{8 \pi}} 
[ \frac{1}{2}I_0(\frac{K}{8 \pi})-\frac{1}{4}I_1(\frac{K}{8 \pi})], \\ 
\nonumber
\frac{\partial y}{\partial l} & = & (2 - \frac{K}{8 \pi})y+2 \pi y^2 
e^{\frac {K}{16 \pi}}I_0(\frac{K}{8 \pi}). 
\label{KTHNY}
\end{eqnarray}
where $I_0$ and $I_1$ are Bessel functions.
The thermodynamic value is recovered by taking the limit $l \to \infty$.

We see in Fig.~(1) that the trajectories in $y$-$K$ plane can be classified
in two classes, namely those for which $y \to 0$ as $l \to \infty$
(ordered phase) and those $y \to \infty$ as $l \to \infty$
(disordered phase).
These two classes of flows are separated by lines called the
separatrix. 
The transition temperature $T_c$ 
(or $\rho_c$) is given by the intersection
of the separatrix with the line of initial conditions $K(\rho,T)$ and
$y = \exp(- E_c(K))$ where $E_c \sim c K/16 \pi$.
At the transition point the flow follows the separatrix
so that the {\em renormalized} $K$ jumps from $16 \pi$ to $0$ at the
transition.
The ordered phase corresponds to the solid (no free dislocations)
and the disordered phase is a phase where free dislocations
proliferate.
Proliferation of dislocations however {\em does not} produce a liquid, rather
a liquid crystalline phase called a ``hexatic'' with quasi- long ranged 
(QLR) orientational order but short ranged positional order.
A {\em second} K-T transition destroys QLR orientational order and takes
the hexatic to the liquid phase by the proliferation of ``disclinations''
(scalar charges).
Apart from $T_c$ there are several universal predictions
from KTHNY- theory, for example, the order parameter correlation length 
and susceptibility has essential singularities ($\sim e^{b t^{-\nu}}, 
t \equiv T/T_c - 1$) near $T_c$.
All these predictions can, in principle, be checked
in simulations\cite{Jas}.

Note that, in order to use the KTHNY- theory to study the solid- hexatic 
transition in hard disks we have to bear in mind that for the 
hard disk solid, which is  
always under a uniform hydrostatic
pressure $P(\rho)$, the effective shear modulus $\mu_{eff}$ has 
to be used in the definition\cite{bates} of $K$.  

The KTHNY- theory predicts when a 2-D solid becomes unstable
to the proliferation of dislocations.
However, there is a second possibility. The free
energy of the liquid may become higher than that of the stable solid 
at a density smaller than that where the hexatic phase is recorded. This
leads to a first order transition and a jump in density at the 
liquid~-solid coexistence pressure (for simulations in the $NVT$- ensemble) 
instead of an intermediate hexatic phase.
Often it is very difficult to distinguish the two
possibilities as the history of simulation studies
of hard disks shows. This is further complicated by the 
fact that KTHNY theory also predicts that the specific heat,
or equivalently, in the case of the hard disk system, the 
compressibility, shows a smooth bump leading to a near flat 
region in the pressure~-density diagram. 
In Fig.~(2a)  we show the conventional situation
where the dotted line designates the often
observed first order transition. In Fig.~(2b) we show Jaster's
results where it is seen that instead of a flat region
in the $P$-$\rho$- curve or a Maxwell loop usually associated
with a first order transition one gets instead a 
smooth bending over to a state with a high compressibility.
Finite size effects which would be present in the first- order
case are negligible.
This would indicate the presence of a liquid- hexatic transition.
The question of solid to hexatic transition is still open. 
It is worth noting that detailed finite size scaling of
orientational order in this system~\cite{henning,mitus}
is not necessarily in contradiction to this result.

Why do simulations of hard disk solids
find it so difficult to see a solid- hexatic transition?
One of the reasons is, of course, the divergence of the
correlation length as the system approaches the transition
so that one requires large systems.
This is complicated by the fact that in order to obtain
equilibrated values of the dislocation density ($\propto y$)
one also needs very large simulation times because
in a high density solid the diffusion of defects is very slow\cite{dmelt}.
To illustrate this point we have attempted to calculate
the defect density of a hard disk solid in a Monte Carlo 
simulation. We perform conventional 
Monte Carlo simulations in the NVT ensemble with an usual Metropolis updating 
scheme for $N=3120$ particles. We choose a single density 
$\rho = .92$; a sequence of initial states are then constructed by 
adding extra complete rows of atoms (thereby increasing the 
density to $\rho_i \geq .92$) and removing an equal number of 
atoms from the bulk at random. In equilibrium, these extra vacancies 
in the bulk should diffuse out and the lattice parameter adjust 
to fill in the gap. After about one million Monte Carlo steps 
we calculate the number of five coordinated ($n_5$) and seven 
coordinated ($n_7$) atoms. Since our system cannot have free vacancies
(due to our choice of ensemble) we expect in equilibrium $n_5 = n_7$.  
The simulations at each $\rho_i$ is 
repeated for ten realizations of the initial state. Our results 
are shown in Fig. (3). We see that $n_5 \neq n_7$ (except for the 
trivial case of $\rho_i = .92$), the difference 
growing with $\rho_i$ as expected, and the statistical errors are 
very large. We therefore conclude that 
even for a relatively small system of $3120$ 
particles, the equilibration of defects (vacancies in this case) 
is an extremely slow process. So it should not come as a surprise 
that brute force simulations of the hard disk solid fail to produce 
the true equilibrium phase.

It may also happen, on the other hand, that KTHNY- theory fails due 
to the following reasons.
Firstly, elastic theory itself may fail near the transition,
so that amplitude or long wave length phonon fluctuations
may destabilise the solid producing a continuous 
transition. Though remote, this possibility has nevertheless
been discussed in the literature~\cite{coments}.
Secondly, perturbation theory in $y$ may break down
because $E_c$ is too small (i.e. $y$ too large) at the transition.
Saito~\cite{saito} and Strandburg~\cite{ecstrand}
showed using lattice discretized versions of a
dislocation Hamiltonian that KTHNY perturbation theory
breaks down if $E_c < 2.7$ at the transition.
In our simulations of the hard disk system we check {\em both}
these possibilities as well as the possibility of a
first order transition.

In the next section we discuss our simulations 
together with  our method for computing elastic constants and
core energies. We use these inputs to check for a first order
transition and a KTHNY- scenario in Section~III.
We summarise and conclude this work in Section~IV.

\section{Elastic constants and 
dislocation core energies from constrained simulations}

One way to circumvent the problem of large finite size effects and 
slow relaxation due to diverging correlation lengths is to simulate 
a system which is constrained to remain defect (dislocation) free and, 
as it turns out, without a  phase transition.
Relatively small systems simulated for short times therefore yields 
thermodynamically accurate data in this limit. Surprisingly, we show that 
using this data it is possible to predict the expected equilibrium behaviour
of the unconstrained system. It is worth mentioning that with 
an approach similar in spirit to the one followed here, we have obtained
excellent results for the Kosterlitz~-Thouless transition in the 
two~-dimensional planar rotor model\cite{EURO},
which has served as an important model in the development
of the KTHNY theory~\cite{KTHNY}, after the proofs of the low temperature
susceptibility divergence in this model~\cite{wegner1} and the
existence of phase transitions without local order parameters
in general~\cite{wegner2} were given.

We simulate $N = 3120$ hard disks in an (almost) square box. We have also 
simulated two additional systems of $N = 2016$ and $N = 4012$ particles 
in order to look for residual finite size effects. Our algorithm follows
closely the usual Metropolis scheme for simulating hard disks. The 
simulation is always started from a perfect triangular lattice which 
fits into our box -- the size of the box determining the density. Once a 
regular MC move is about to be accepted, we perform a  
{\em local} Delaunay triangulation involving the moved disk and its 
nearest and next nearest neighbors. We compare the connectivity of 
this Delaunay triangulation with that of the reference lattice (a copy 
of the initial state) around the same particle. If any old bond is 
broken and a new bond formed (Fig. (4)) we reject the move since one 
can show that this is equivalent to a dislocation~- antidislocation pair 
separated by one lattice constant involving dislocations of the smallest
Burger's vector. Note that, (i) only  dislocation pairs of smallest Burger's 
vector are eliminated, dislocations of higher topological charge cost higher
energy and may not be relevant at the densities where a melting transition
is usually observed; (ii) other fluctuations e.g. long wavelength phonon 
fluctuations and fluctuations of the amplitude of the order parameter 
(spontaneous production of voids in the system) are not eliminated as long
as they preserve connectivity. The fraction of moves $p$ which are rejected 
because they violate the constraint is stored. Next, we 
need a method to calculate elastic constants accurately in our simulations 
making sure that we extrapolate to the thermodynamic limit. Such a 
method has been recently developed by us and discussed in detail 
elsewhere\cite{SNRB}. Below we include a brief description for completeness.

Since we have a dislocation free system, we can always associate an ideal, 
static, ``reference'' lattice point ${\bf R}$ with every hard disk all 
through the simulation and calculate ${\bf u}_{\bf R}(t)={\bf R}(t)-{\bf R}$. 
Microscopic strains $\epsilon_{ij}({\bf R})$ can be calculated now for every 
reference lattice point ${\bf R}$. Next, we  
coarse grain (average) the microscopic strains within a 
sub-box of size $L_b$,
$$ \bar{\epsilon}_{ij} = L_b^{-d} \int^{L_b} d^dr \epsilon_{ij}({\bf r}) $$
and calculate the ($L_b$ dependent) quantities,
\begin{eqnarray}
S_{++}^{L_b} & = & <\bar{\epsilon}_{+}\bar{\epsilon}_{+}> \\ \nonumber
S_{--}^{L_b} & = & <\bar{\epsilon}_{-}\bar{\epsilon}_{-}> \\ \nonumber
S_{33}^{L_b} & = & 4<\bar{\epsilon}_{xy}\bar{\epsilon}_{xy}>  
\label{sgam}
\end{eqnarray}

\noindent
The sub-blocks may be constructed by simply dividing the entire box of 
size $L$  
into integral number of smaller boxes, as done in this calculation 
so that $L/L_b = $ an integer, or multiple sub-boxes of arbitrary size 
$L_b \leq L$ can be constructed within the simulation cell, as in 
Ref.\cite{SNRB}.
Lastly, quantities in the thermodynamic limit are obtained by fitting data 
to the form,

$$ S_{\gamma\gamma}^{L_b} = S_{\gamma\gamma}^\infty 
\left[ \Psi(x L/\xi)-\left
(\Psi(L/\xi)-C\left(\frac{a}{L}\right)^2\right) x^2 \right] 
                       + {\cal O}(x^4). $$ 
where the index $\gamma = +, - , 3$ the function, $\Psi(\alpha)$ is defined as,
$$
\Psi(\alpha) = \frac{2}{\pi}\alpha^2 \int_0^1 \int_0^1 dx dy\,\, K_0 (\alpha \sqrt{x^2+y^2})
$$
$K_0$ is a Bessel function and $\xi$ is the correlation length for 
the $\epsilon\epsilon$- correlations.

The elastic constants in the thermodynamic limit are obtained from,
the set: $B = 1/2 S_{++}^{\infty}$ and 
$\mu_{eff} = 1/2 S_{--}^{\infty} = 1/2 S_{33}^{\infty}$. 
The last
two equations for $\mu_{eff}$ serve as a stringent internal consistency 
check and yields an accurate error estimate for this quantity. There 
are two ways to obtain the fluctuations $ S_{\gamma\gamma}^{L_b}$ for 
every sub-block size $L_b$ in Eq. (5). One can either accumulate 
$<\epsilon_{\gamma}\epsilon_{\gamma}>$ directly or construct histograms 
of the block strains $\epsilon_{\gamma}$ and obtain $S_{\gamma\gamma}$ 
by fitting Gaussian profiles to the normalized probability distributions of 
$\epsilon_{\gamma}$ for every block size $L_b$. Again this constitutes 
another excellent consistency check and a measure of the statistical 
uncertainties involved. We accumulate data till all these uncertainties
are less than a percent. Residual finite size effects obtained by 
repeating the entire procedure for $N= 2016$ and $4012$ particles for 
a few densities are also seen to be within the same limit of accuracy. 

There are several distinct advantages of our method:
In general our method works for any system for which instantaneous 
configurations can be obtained (for example either from other simulations or
from real experiments). We obtain directly the finite size scaled 
results from a single simulation. As discussed above there are
a number of stringent internal consistency checks which
can be used to obtain very accurate data.
In spite of this our method is easy to use and the
computational complexity is not more than calculating
for eg. pair correlation functions.
This method can be easily adapted for calculating 
local strains and stresses in {\em inhomogeneous} situations.

\section{Results and Discussion}

Our results for the elastic moduli, the pressure and the fraction 
of moves, $p$, rejected due to the topological constraint discussed above 
are given in Table I as a function of density. In Fig.~(5), we compare 
our results for the bulk and shear moduli with the data of two previous
simulations of Ref.~\cite{WB} and Ref.~\cite{blad}.
We also compare our simulation results to estimates from free volume 
theory\cite{frevol} in the simplest, independent cell approximation. Within 
this approach the Helmholtz free energy per particle is given by 
$ f = \log(v_f)$, where the available free volume, 
$v_f = (a-1)^2/\rho_c$ and the close packed density $\rho_c = 2/\sqrt{3}$.
Other thermodynamic quantities can be obtained by successive 
differentiation, viz. 
\begin{eqnarray}
P & = & \rho \frac{x}{x-1} \\ \nonumber
B & = & P \lbrace 1 + \frac {1}{2 (x-1)} \rbrace \\ \nonumber
\mu_{eff} & = & \frac{B}{2}
\end{eqnarray}
where $x = \sqrt{\rho_c/\rho}$ and we have used the Cauchy relation,
strictly valid only for a harmonic solid\cite{SNRB}, for our estimate of the 
effective shear modulus $\mu_{eff}$. Note that the free volume elastic moduli 
and the pressure diverge\cite{frevol} as $\rho \to \rho_c$.  

We see that our bulk modulus interpolates smoothly
from the free volume values
at high densities to those of Ref.~\cite{blad} at low densities.
Overall, the differences between the three sets of data are small.
Our values for the shear modulus agrees well with the
free volume results at high density, but at low densities
they are smaller than all other estimates
though close to those of Ref.~\cite{WB}. 

Once the elastic constants are obtained
we can analyze in detail the two competing scenarios
viz. first order solid- liquid transition or
KTHNY- transition to the hexatic phase.

\subsection{Equation of state, free energy and first order melting}

First of all, we should point out that our constrained 
simulations allow us to obtain elastic constants up to a density
as low as $\rho=.88$, far below the density $\rho = .899$ where the 
transition to the liquid is expected to occur\cite{Jas}, which implies 
that amplitude and phonon
fluctuations cannot destabilize the solid. So an ordinary
second order transition is ruled out. However,
there can always be a first order transition if the free energy
of the liquid becomes lower than that of the perfect solid.

In order to investigate this question we obtain the equation of state
$P(\rho)$ and the Gibbs free energy $g(P)$ of the liquid
and the solid.

To obtain the equation of state of the liquid 
we use the semi~-empirical, accurate, analytical form by 
Santos {\em et. al.}~\cite{eos2}, which is in excellent agreement 
with computer simulation data\cite{eos1}. The pressure is given by,
\begin{equation}
P/\rho = \lbrace 1 -2 \eta +\frac{2 \eta_c - 1}{\eta_c^2}\eta^2 \rbrace^{-1}
\end{equation}
where the packing fraction $\eta = (\pi/4) \rho$ and $\eta_c$ is the 
packing fraction at close packing. 
The Helmholtz free energy per particle, 
\begin{equation}
f(\rho) = \int_0^\eta d \eta' \frac{P/\rho -1}{\eta'} + f_{id}, 
\label{santos}
\end{equation}
where the ideal gas Helmholtz free energy per particle 
$f_{id} = \log(\rho)-1$. The Gibbs free energy $g(P)$ is then obtained by 
the standard Legendre transformation, $g = f + P/\rho$.
In addition we use the data of Jaster~\cite{Jas} in 
the transition region to obtain a revised estimate of the free energy.
This is done by fitting Jaster's data for $P(\rho)$ 
to a polynomial for $\rho > 0.85$ which matches the results of 
Santos {\em et. al.}~\cite{eos2} for $\rho \leq 0.85$.
From this equation of state we can obtain the Helmholtz and hence the
Gibbs free energy by integrating starting from the value given by 
Eq.(7) at $\rho=0.85$.

The equation of state for the solid is obtained by integrating our bulk
modulus values using the result of Bladon and Frenkel~\cite{blad} at 
$\rho = 1.049$ as the reference pressure ($P = 22.00$).
The Gibbs free energy is obtained by further integration
again using the result obtained for the free energy in Ref.~\cite{blad}
at $\rho = 1.049$ as a reference ($g = 25.64$).

The possible (first order) transitions can be located by equating the 
Gibbs free energies. The slope discontinuity gives the (inverse) density 
difference of coexisting phases. We find immediately, that all the free 
energies have very similar slopes (see Fig. 6) so that any possible 
first order transition would have only a small jump in the density. It 
also implies that small errors in the free energy of our reference state
makes a large difference in the co-existence pressure. We have therefore
reduced the reference free energy by a small amount ($ < 4 \%$) so that 
the coexisting pressure $P_1 = 9.2$ --- the value found in most 
recent simulations\cite{zolche,Jas}. 
 
Using the semi-empirical free energy of Santos {\em et. al.}~\cite{eos2}
we obtain a (meta stable) first order transition with $\rho_l = 0.871$ and 
$\rho_s = 0.912$ as observed in early simulations\cite{aldwain,wwwood}.
Of course, this estimate of $\rho_l$ is only a lower bound, as the 
theory of Ref.\cite{eos2} is expected to overestimate the free energy.
The free energy from Jaster's data is lower and almost completely parallel
to that of the solid suggesting a very weak first order transition if at all.
In this case we get a slope difference $< 1.3 \%$ 
(viz. $\rho_l = 0.899$ and $\rho_s = 0.911$) - well 
within our numerical accuracy (Fig. 6).

\subsection{Core energy $E_c$ and the KTHNY transition:}

Next, we analyze  our results in the light of the KTHNY- theory\cite{KTHNY}.
The {\em unrenormalized}  $K = 16 \pi$ at 
$\rho_c = 0.904$ ($P_c = 8.92$) (see Fig.~(2), lower arrow)
which implies that a weak
first order transition from solid to liquid preempts a KTHNY- solid- hexatic
transition. However, the value of $K$ is renormalized by the presence
of dislocations. We can estimate the extent of this renormalization
from our data.  

The dislocation pair probability 
\begin{equation}
p_d = \exp(-2E_c)Z(K)
\end{equation} 
where $Z(K)$ is
the ``internal partition function'' of a dislocation pair and is given 
by\cite{morf},
\begin{equation}
Z(K) = \frac{2 \pi \sqrt{3}}{K/8 \pi -1}I_0(\frac{K}{8\pi})
\exp(\frac{K}{8\pi}).
\end{equation}
Where we have set the core radius $r_c = a$, the lattice parameter.
The core energy of a dislocation is a  difficult quantity to obtain 
from a simulation, though it has been attempted in the past\cite{blad,morf}.
In our case,
an ansatz, which gives excellent results in the 2D- XY- model~\cite{EURO},
and identifies the rejection ratio $p$ as $p=p_d$ can be used to obtain 
$E_c$, see Fig.~(7).
Throughout the relevant region $E_c$ is
safely above the limit $E_c > 2.7$~\cite{ecstrand,saito}. 
At the transition the $E_c \sim 6$ which is in good agreement
the results of Murray and Van Winkle~\cite{MW} ($E_c \sim 5.6$)
from experiments on 2-d charge stabilised colloids 
and of Zahn {\em et. al.}~\cite{maret} ($E_c \sim 4.$) for paramagnetic 
colloids. 

Finally, to obtain the melting density we use the unrenormalised $K$ and 
$y = \exp(- E_c(K))$ as inputs to the KTHNY recursion relations 
(Eqs.(4)) and solve them numerically by a standard
Euler discretization to obtain $K_R$, see Fig.~(8).
The melting density obtained from our value for $K_R$ is
$\rho_c = .916$ and $P_c = 9.39$ 
(Fig.~(2), upper arrow).
This means that the KTHNY- transition now preceeds the first order
transition and the solid transforms to the hexatic phase.

\section{Summary and conclusion}
We have simulated a dislocation free triangular solid of hard disks using 
a constrained Monte Carlo algorithm. Using a block analysis 
scheme we calculate the finite size scaled elastic constants of this 
solid. From the number of times the system attempts to violate our no-
dislocation constraint we can obtain (virtual) dislocation probabilities
and hence the core energy. The absence of a phase transitions in our system 
implies that all correlation lengths remain finite and the problem of 
slow equilibration of defect densities is eliminated. In effect we 
obtain highly accurate values of the unrenormalized coupling constant
$K$ and the defect fugacity $y$ which can be used as inputs to the KTHNY 
recursion relations. Numerical solution of these recursion relations then 
yields the renormalized coupling $K_R$ and hence the density and pressure 
of the solid to hexatic melting transition.

We can draw a few very precise conclusions from our results.
Firstly, a solid without dislocations is stable against fluctuations
of the amplitude of the solid order parameter and against long wavelength 
phonons. So any melting transition mediated by phonon or 
amplitude fluctuation is ruled out in our system. Secondly,  
the core energy $ E_c  > 2.7$ at the transition so KTHNY
perturbation theory is valid though numerical values of nonuniversal 
quantities may depend on the order of the perturbation analysis. 
Thirdly, solution of the recursion relations shows that  
a KTHNY transition at $P_c = 9.39$ {\em preempts}
the first order transition at $P_{1} = 9.2$. Since these transitions,
as well as the hexatic~-liquid KTHNY transition lies so close to each other,
the effect of, as yet unknown, higher order corrections to the recursion 
relations may need to be examined in the future\cite{EURO}. Due to this
caveat, our conclusion that a hexatic phase exists over some region of
density exceeding $\rho = .899$ still must be taken as preliminary. 
Also, in actual simulations,
cross over effects near the bicritical point, where two critical lines 
corresponding to the liquid~-hexatic and hexatic~-solid transitions meet 
a first order liquid~-solid line (see for e.g. Ref.\cite{jankl} for a
lattice model where such a situation is discussed) may complicate the 
analysis of the data, 
which may, in part, explain the confusion which persists in the literature 
on this subject. In systems with softer potentials\cite{bgw}, the 
signature of a KTHNY transition appears to be more pronounced\cite{r12mel}.
In future, we would like to analyze more complicated systems eg. laser 
induced reentrant melting of charge stabilized colloids\cite{bech}, 
and the influence of other defect variables eg. grain boundaries\cite{GBM}
on elastic constants and melting behaviour.

\section{Acknowledgement}

We are grateful for many illuminating discussions with D. Frenkel, 
M. Bates, Madan Rao, W. Janke and D. R. Nelson. One of us (S.S.) thanks
the Alexander von Humboldt Foundation for a Fellowship.
Support by the SFB~513 is gratefully acknowledged. This paper is 
dedicated to F. J. Wegner on the occasion of his 60$^{th}$ 
birthday.


\newpage
\begin{tabular} {||l|l|l|l|l|l|l||} \hline
$\rho$ & $N_c$ & $P$  & $B$ & $\mu_{eff}$ & $p\times10^2$  &  $K/16 \pi$\\ \hline
  0.88& 10$^5$& 8.117& 27.69&  11.63& 0.36823& 0.8550 \\
  0.9& 10$^5$&  8.777& 32.47&  13.87& 0.20358&  0.9925 \\
  0.905&10$^5$&  8.957&  33.67&   14.46& 0.17386& 1.0271\\
  0.910&10$^5$&  9.145& 35.38&  15.22& 0.14469& 1.0744 \\
  0.915&10$^5$&  9.342& 37.09&  15.99& 0.11706& 1.1225 \\
  0.920&10$^5$&  9.545& 38.48&  16.88& 0.09532& 1.1722 \\
  0.925&10$^5$&  9.759& 40.67&  17.88& 0.07513& 1.2337 \\
  0.930&10$^5$&  9.982& 42.72&  18.90& 0.05967& 1.2948 \\
  0.935&10$^5$&  10.217& 44.69&  19.91& 0.04643& 1.3538 \\
  0.94&2$\times$10$^4$&   10.462& 46.85&   21.45& 0.03432& 1.4382 \\
  0.95&10$^5$&   10.996& 52.14&  24.10&  0.01855& 1.5945 \\
  0.96&2$\times$10$^4$&   11.586& 59.67&  27.61&  0.00901& 1.8067 \\
  0.97&10$^5$&   12.251& 67.45&   31.59& 0.00370& 2.0379 \\
  0.98&2$\times$10$^4$&   13.003& 79.20&  36.62&  0.00137& 2.3479 \\
  0.99&10$^5$&   13.862& 89.98&  42.60&  0.00041& 2.6835 \\
  1.&49400&    14.843& 104.78&   50.25& 0.00009& 3.1206 \\
  1.02&10$^5$&  17.301& 148.88 &  69.91&  0.0& 4.2854 \\
  1.04&10$^5$&   20.714& 212.02&   102.02& 0.0&    6.0857 \\
  1.06&10$^5$&   -& 319.07&   158.69&   0.0& 9.1874 \\
  1.08&10$^5$&   -& 531.24&  268.02&    0.0 &15.1567 \\
  1.1&10$^5$&  -& 1018.49&  526.94&    0.0& 29.0094 \\ \hline
\end{tabular}
\vskip .5cm

\noindent
{\bf Table~I}~~ Pressure $P$, bulk modulus $B$, effective shear modulus 
$\mu_{eff}$,
ratio of moves rejected due to the zero dislocation density 
constraint $p$ and the (unrenormalized) coupling constant $K/16 \pi$ as
a function of the density $\rho$. The total number of configurations 
used for the averages $N_c$ is also listed. The pressure $P$ was 
obtained by integrating $B$ below $\rho = 1.049$.

\newpage
\begin{figure}[hbtp]
\begin{picture}(0,100)
\put(0,0) {\psfig{figure=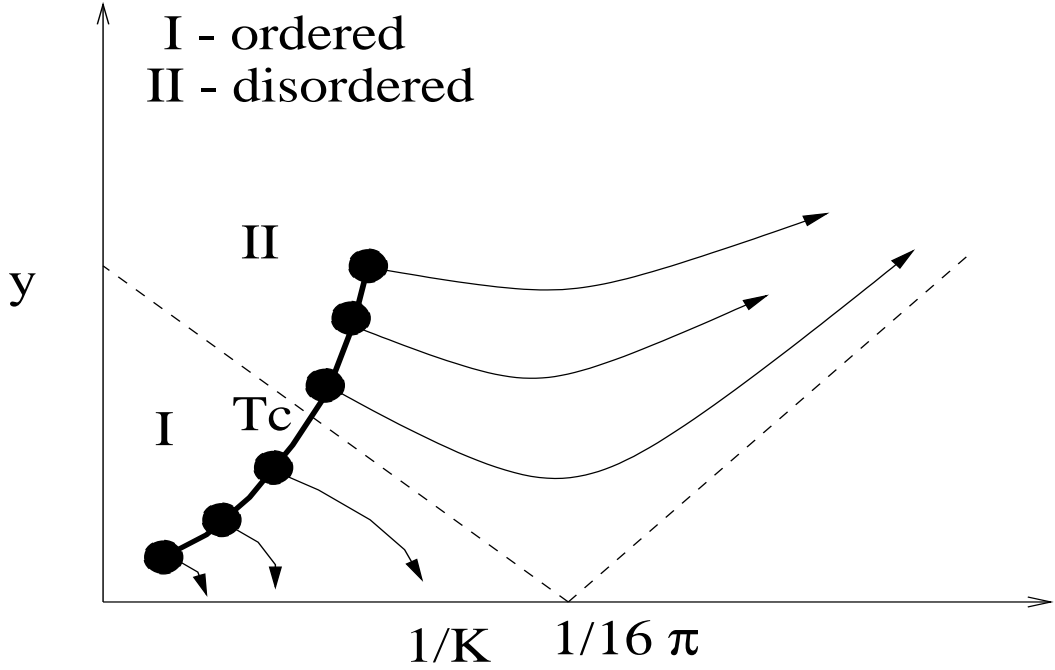,width=139mm,height=88mm}}
\end{picture}
\vskip .5 cm
\caption[]
{Schematic flows of the coupling constant $K$ and the defect 
fugacity $y$ under the action of the KTHNY recursion relations.
The dashed line is the separatrix whose intersection with the 
line of initial state (solid line connecting filled 
circles, $y(T,l=0), K^{-1}(T,l=0)$) determines the transition point $T_c$}
\end{figure}
\newpage

\begin{figure}[hbtp]
\begin{picture}(0,80)
\put(0,0) {\psfig{figure=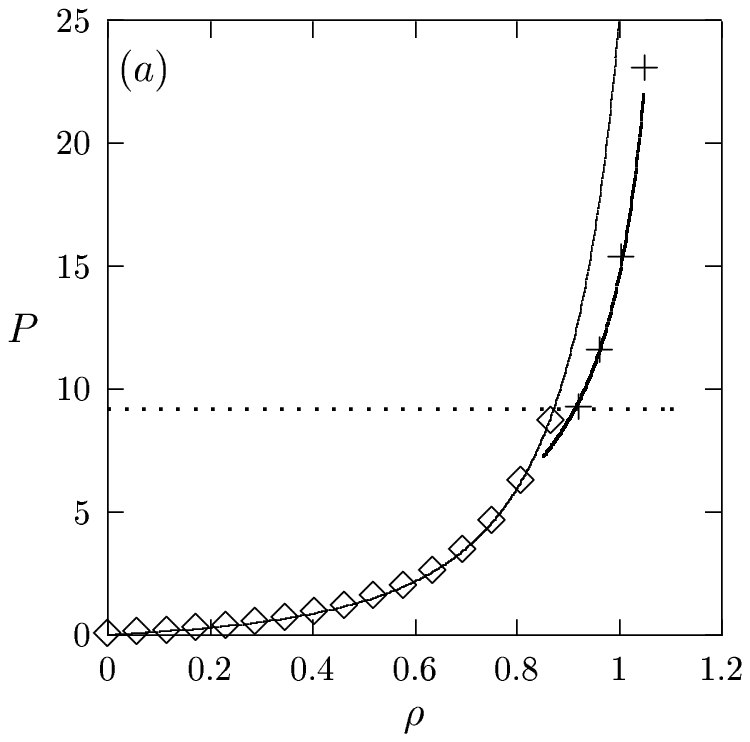,width=70mm,height=70mm}}
\put(70,0) {\psfig{figure=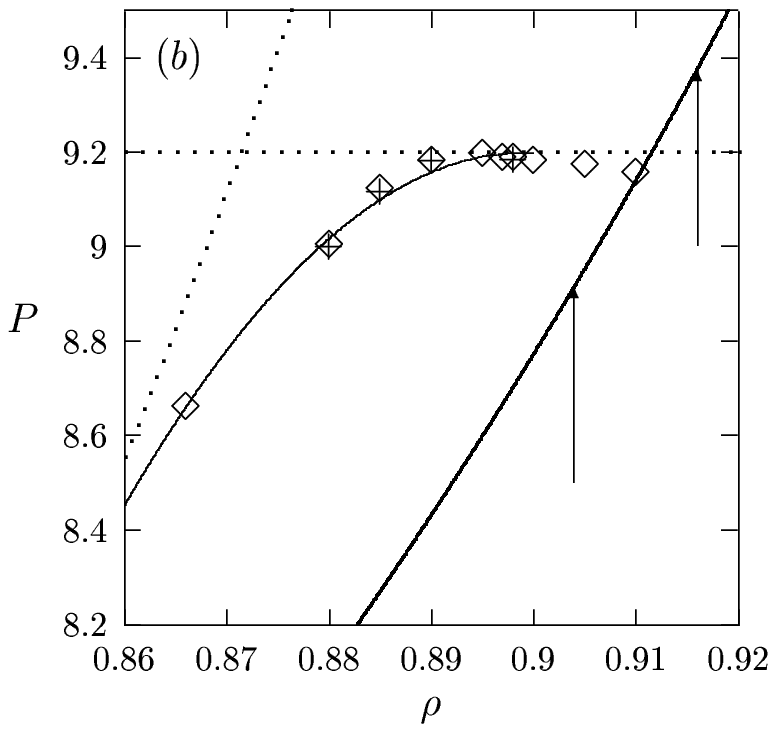,width=70mm,height=70mm}}
\end{picture}
\caption[]
{Equation of state of the hard disk -liquid and -solid.
(a) Liquid: light solid line; semi-empirical form of Santos et al.~\cite{eos2},
$\Diamond$; data from Hoover et al.~\cite{eos1}. Solid: bold solid line;
our results, $+$; data of Wojciechowski and Bra\'nka~\cite{WB}, dotted line;
position of coexistence pressure as seen in all studies observing a first order
phase transition.
(b) Expanded view of (a) near the phase transition region.
$\Diamond$; results  of Jaster~\cite{Jas} for 128 $\times$ 128 particles,
$+$; same for 256 $\times$ 256 particles. Light solid line;
polynomial fit to Jaster's data, bold solid line; our data for the solid
(as in (a)). Horizontal dotted line: as in (a), dotted curve; semi-empirical 
form for the equation of state of the hard disk liquid of Santos 
{\em et. al.}\cite{eos2}. Arrows: lower arrow; position of the
KTHNY-transition with bare values for $K$, upper arrow;
same with renormalized $K_R$ calculated from our simulations.
Note that the accuracy of Jaster's data is smaller than the size of
the symbols for $\rho \leq .9$, while for $\rho > .9$ there may be 
systematic finite size effects and finite observation time effects 
possibly invalidating the data.
}
\end{figure}
\newpage

\begin{figure}[hbtp]
\begin{picture}(0,100)
\put(0,0) {\psfig{figure=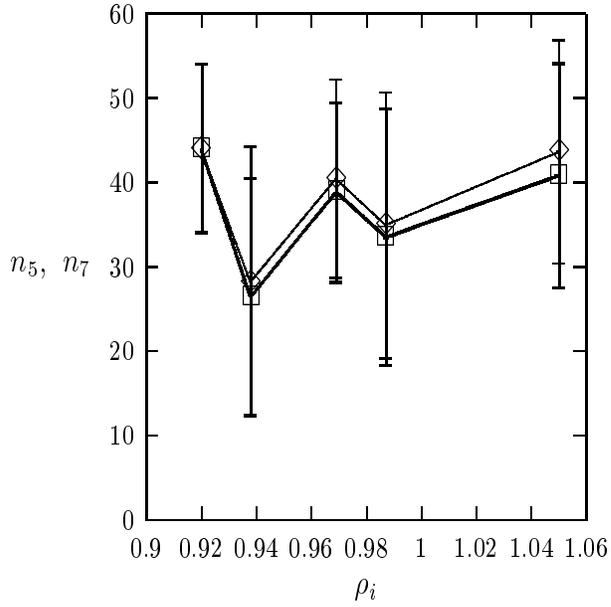,width=80mm,height=80mm}}
\end{picture}
\vskip .5 cm
\caption[]
{The number of hard disks with five fold ($n_5$, $\Diamond$ and light 
 solid line) and seven fold ($n_7$, $+$ and bold solid line) 
co-ordination after $10^6$ Monte Carlo steps per particle for a 
$N = 3120$ particle system, plotted against $\rho_i$ (see text). Note 
that $n_5 \ne n_7$ for $\rho_i$  larger than $.92$. 
}
\end{figure}
\newpage

\begin{figure}[hbtp]
\begin{picture}(0,80)
\put(0,0) {\psfig{figure=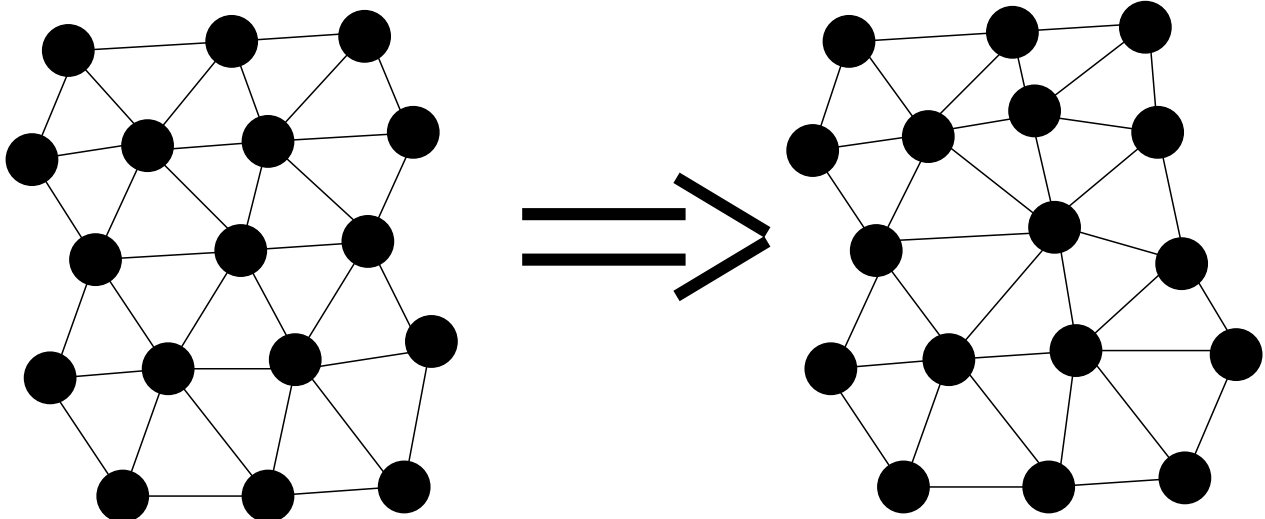,width=167.755mm,height=70mm}}
\end{picture}
\caption[]
{Typical move which attempts to change the coordination number
and therefore the local connectivity around the central particle.
Such moves were rejected in our simulation.}
\end{figure}
\newpage

\begin{figure}[hbtp]
\begin{picture}(0,80)
\put(00,0) {\psfig{figure=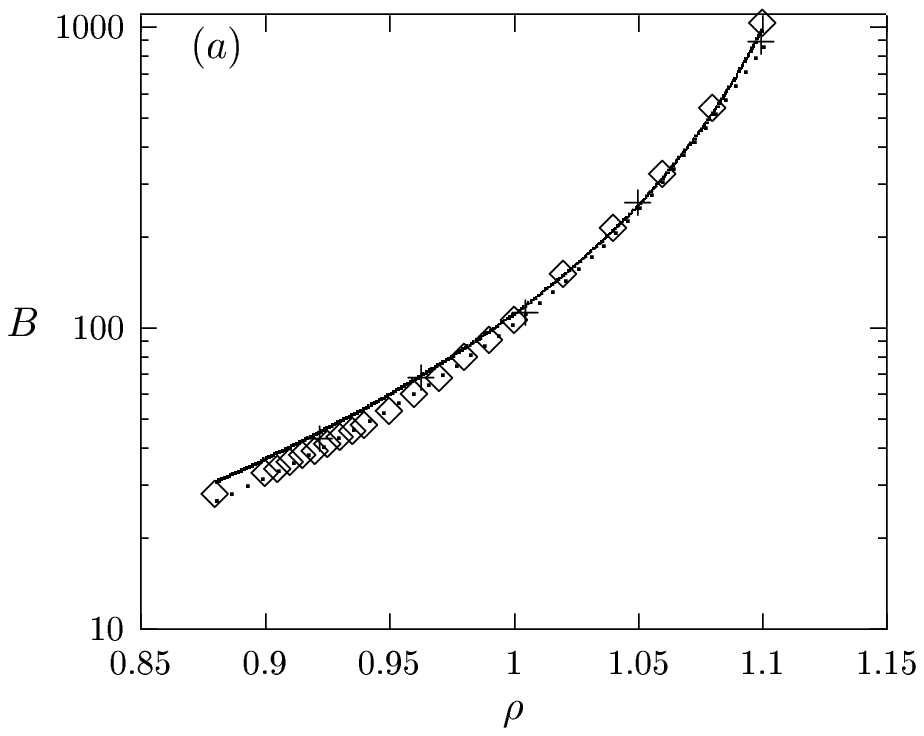,width=70mm,height=70mm}}
\put(70,0) {\psfig{figure=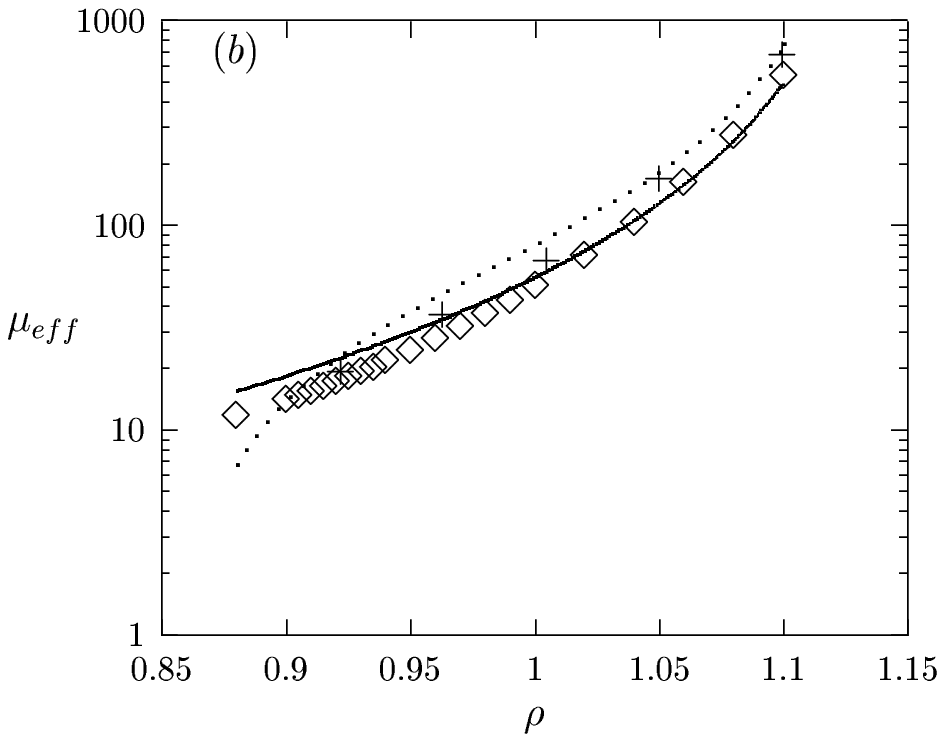,width=70mm,height=70mm}}
\end{picture}
\caption[]
{
Elastic moduli in the thermodynamic limit:
(a) bulk $B$ and (b) shear $\mu_{eff}$.
$\Diamond$ - our work (error bars are much smaller than the symbol size);
$+$ Wojciechowski and Bra\'nka~\cite{WB}, solid line;
free volume theory~\cite{frevol}, dashed line; 
polynomial fit given by P. Bladon and D. Frenkel~\cite{blad}. 
}
\end{figure}
\newpage

\begin{figure}[hbtp]
\begin{picture}(0,100)
\put(50,0) {\psfig{figure=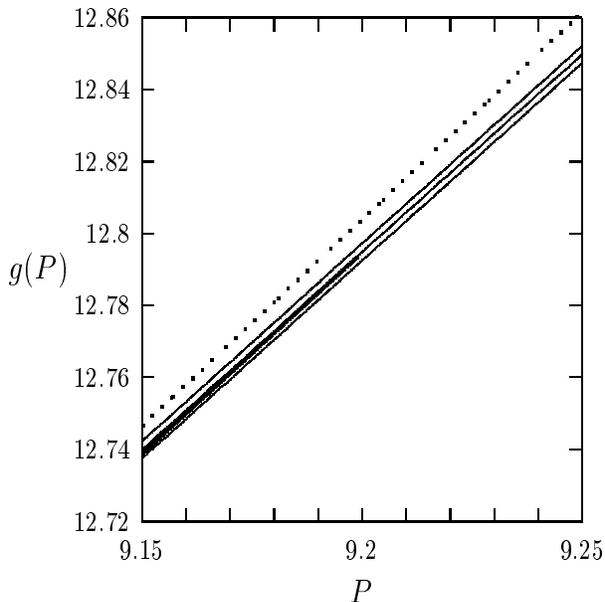,width=80mm,height=80mm}}
\end{picture}
\caption[]
{
Gibbs Free energies $g(P)$ as a function of pressure:
dotted line; metastable liquid using semi empirical form of Santos
{\em et. al.}~\cite{eos2};
bold solid line; using Jaster's results~\cite{Jas},
series of light solid lines; gibbs free energy of the solid where we 
reduced the reference free energy from 
the value quoted by Bladon and Frenkel~\cite{blad} 
(see text) by $3.3, 3.35$ and $3.4 \%$. 
}
\end{figure}
\newpage

\begin{figure}[hbtp]
\begin{picture}(0,80)
\put(50,0) {\psfig{figure=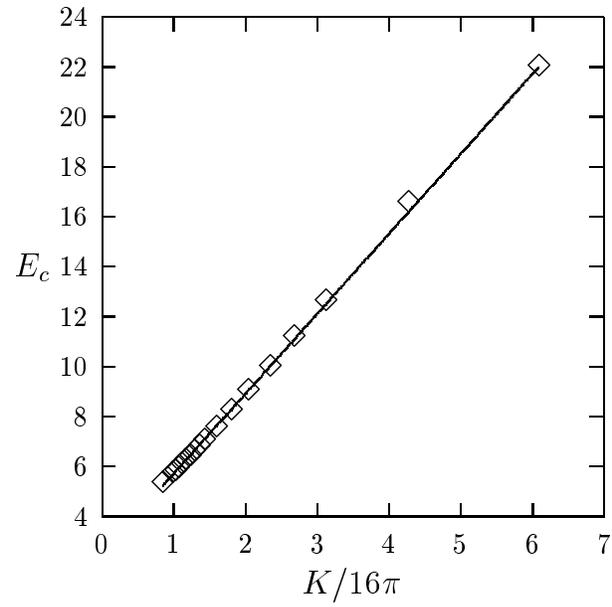,width=80mm,height=80mm}}
\end{picture}
\caption[]
{Calculated core energy $E_c$ ($\Diamond$) as a function of $K/16\pi$.
The straight line is a linear least square fit.
Note that $E_c > 2.7$ throughout.
}
\end{figure}
\newpage

\begin{figure}[hbtp]
\begin{picture}(0,100)
\put(20,0) {\psfig{figure=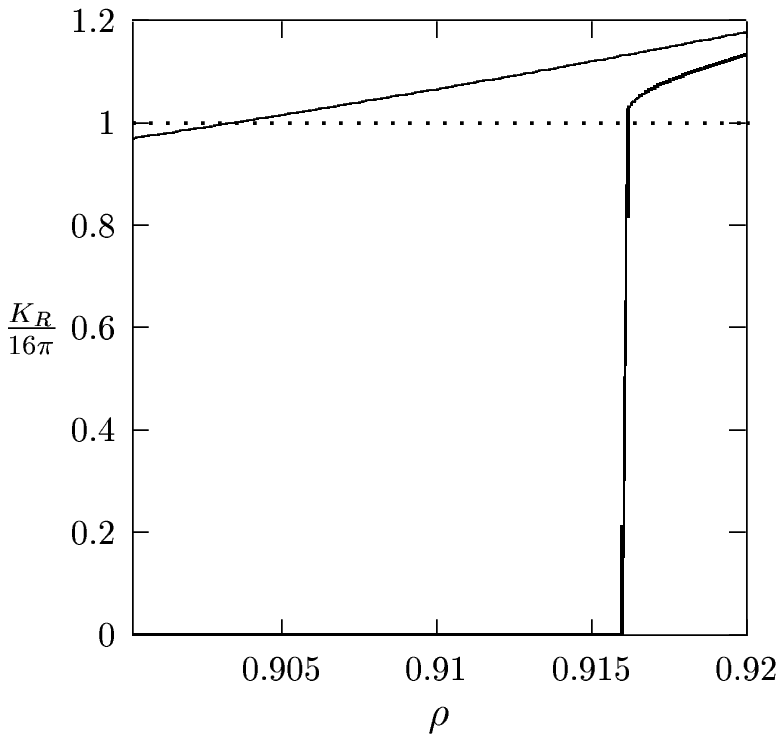,width=80mm,height=80mm}}
\end{picture}
\caption[]
{
Renormalization of $K/16 \pi$ vs density $\rho$ for the hard disk solid.
The renormalized $K_R/16 \pi$ (bold solid line)i is obtained from the 
recursion relations Eq. (4) which were solved by Euler discretization 
using a step size $\delta l = 0.001$ upto a final $l = 100$ starting from 
the initial input (light solid line).
Dotted line: $K =16 \pi$.
}
\end{figure}


\begin{references}
\bibitem{metro} N. Metropolis, A. W. Rosenbluth, M. N. Rosenbluth, 
A. H. Teller, E. Teller, J. Chem. Phys. {\bf 21}, 1087 (1953).

\bibitem{aldwain} B. J. Alder and T. E. Wainwright, Phys. Rev. {\bf 127}, 359
(1962).

\bibitem{simple} The simplicity of the hard disk potential allows for 
extremely efficient simulation algorithms. See for example, M. Isobe,
Int. J. Mod. Phys. C, {\bf 10}, 1281 (1999), for a recent ${\cal O}(\log N)$
algorithm for hard disk molecular dynamics.

\bibitem{wwwood} W. W. Wood in {\em Physics of Simple Liquids}, edited by 
H. N. V. Temperley, J. S. Rowlinson and G. S. Rushbrooke (North-Holland, 
Amsterdam, 1968), Chap. 5.

\bibitem{hdstrand} J. Lee and K. Strandburg, \prb {\bf 46}, 11190 (1992).

\bibitem{zolche} J. A. Zollweg and G. V. Chester, \prb {\bf 46}, 11186 (1992).

\bibitem{dft} T. V. Ramakrishnan, Phys. Rev. Lett. {\bf 42}, 795 (1979);
X. C. Zeng and D. W. Oxtoby, J. Chem. Phys. {\bf 93}, 2692
(1990); Y. Rosenfeld, \pra, {\bf 42}, 5978 (1990).
V. N. Ryzhov and E. E. Tareyeva, \prb, {\bf 51}, 8789 (1995).

\bibitem{Jas} A. Jaster, Phys. Rev. E {\bf 59}, 2594 (1999).

\bibitem{KTHNY} 
J. M. Kosterlitz, D. J. Thouless, J. Phys. {\bf C 6}, 1181 (1973); 
B.I. Halperin and D.R. Nelson, \prl {\bf 41}, 121 (1978);
D. R. Nelson and B. I. Halperin, Phys. Rev. B {\bf 19}, 
2457 (1979);
A.P. Young,
Phys. Rev. B {\bf 19}, 1855 (1979); K.J. Strandburg, Rev. Mod. Phys. {\bf 60},
161 (1988); H. Kleinert, {\em Gauge Fields in Condensed Matter}, Singapore,
World Scientific (1989).

\bibitem{WB} K. W. Wojciechowski and A. C. Bra\'nka, Phys. Lett. {\bf 134A},
314 (1988).

\bibitem{bates} M. Bates and D. Frenkel, preprint.

\bibitem{henning} H. Weber, D. Marx and K. Binder, \prb {\bf 51}, 14636 (1995);
H. Weber and D. Marx, Europhys. Lett. {\bf 27}, 593 (1994).

\bibitem{mitus} A. C. Mitus, H. Weber and D. Marx, \pre {\bf 55}, 6855 (1997).

\bibitem{dmelt} A. Zippelius, B. I. Halperin and D. R. Nelson, \prb, {\bf 22},
2514, (1980).

\bibitem{coments} J. F. Fern\'andez, J.J. Alonso and J. Stankiewicz, \prl
{\bf 75}, 3477; H. Weber and D. Marx, \prl {\bf 78} 398, (1997); 
J. F. Fern\'andez, J.J. Alonso and J. Stankiewicz, \prl {\bf 78}, 399 (1997).

\bibitem{saito} Y. Saito, \prl {\bf 48}, 1114 (1982); \prb {\bf 26}, 6239 
(1982).

\bibitem{ecstrand} K. J. Strandburg, \prb {\bf 34}, 3536 (1986).

\bibitem{EURO} S. Sengupta, P. Nielaba, K. Binder,
preprint, cond-mat/0001309 

\bibitem{wegner1} F. J. Wegner, Z. Phys. {\bf 206}, 465 (1967).

\bibitem{wegner2} F. Wegner, J. Math. Phys. {\bf 12}, 2259 (1971).

\bibitem{SNRB} S. Sengupta, P. Nielaba, M. Rao and K. Binder cond-mat/9906063 
(to appear in Phys. Rev. E).

\bibitem{frevol} W. G. Hoover, W. T. Ashurst and R. Grover, J. Chem. Phys. 
{\bf 57}, 1259 (1972); W. G. Hoover, N. E. Hoover and K. Hanson, J. Chem. Phys.
{\bf 70}, 1837 (1979);
W. G. Hoover and F. H. Ree, J. Chem. Phys. {\bf 49}, 3609
(1968).

\bibitem{eos2} A. Santos, M. L\'opez de Haro and S. Bravo Yuste, J. Chem. Phys.
{\bf 103}, 4622 (1995).

\bibitem{eos1} B. J. Alder, W. G. Hoover and D. A. Young, J. Chem. Phys.
{\bf 49}, 3688 (1968).

\bibitem{blad} P. Bladon and D. Frenkel, preprint (2000).

\bibitem{morf} D. S. Fisher, B. I. Halperin and R. Morf, \prb {\bf 20}, 4692
(1979).

\bibitem{MW} C. A. Murray and D. H. Van Winkle, Phys. Rev. Lett. {\bf 58},
1200 (1987).

\bibitem{maret} K. Zahn, R. Lenke and G. Maret, \prl {\bf 82}, 2721, (1999)

\bibitem{jankl} W. Janke and H. Kleinert, \prl {\bf 61}, 2344 (1988).

\bibitem{bgw} J. Q. Broughton, G. H. Gilmer and J. D. Weeks, \prb,
{\bf 25}, 4651 (1982).

\bibitem{r12mel} K. Bagchi, H. C. Andersen and W. Swope, \pre {\bf 53},
3794 (1996).

\bibitem{bech} Q.~-H. Wei, C. Bechinger, D. Rudhardt and P. Leiderer, \prl
{\bf 81}, 2606 (1998).

\bibitem{GBM} S.T. Chui, Phys. Rev. Lett. {\bf 48}, 933 (1982); \prb {\bf 28},
178 (1983).
\end{references}
\end{document}